\documentclass[twocolumn,showpacs,preprintnumbers,amsmath,amssymb]{revtex4}
\usepackage{graphicx}
\usepackage{dcolumn}
\usepackage{bm}
\usepackage{color}
\usepackage{rotating}

\begin{document}
\title{Optimizing the Multi-Photon Absorption Properties of N00N States}
\author{William~N.~Plick, Christoph~F.~Wildfeuer, Petr~M.~Anisimov, and Jonathan~P.~Dowling}
\affiliation{Hearne Institute for Theoretical Physics, Department of Physics and Astronomy, Louisiana State University, Baton Rouge, LA 70803.}
\date{\today}

\begin{abstract}
In this paper we examine the $N$-photon absorption properties of ``N00N" states, a subclass of path entangled number states. We consider two cases. The first involves the $N$-photon absorption properties of the ideal N00N state, one that does not include spectral information. We study how the $N$-photon absorption probability of this state scales with $N$. We compare this to the absorption probability of various other states. The second case is that of two-photon absorption for an $N=2$ N00N state generated from a type II spontaneous down conversion event. In this situation we find that the absorption probability is both better than analogous coherent light (due to frequency entanglement) and highly dependent on the optical setup. We show that the poor production rates of quantum states of light may be partially mitigated by adjusting the spectral parameters to improve their two-photon absorption rates. This work has application to quantum imaging, particularly quantum lithography, where the N-photon absorbing process in the lithographic resist must be optimized for practical applications.
\end{abstract}

\maketitle

\section{Introduction}
We investigate the multiphoton
absorption probabilities of maximally path entangled number states
\--- also called $N00N$ states, after the way the state vector is
written: $|N::0\rangle\equiv(|N,0\rangle+|0,N\rangle)/\sqrt{2}$. The two positions in each state vector represent two spatial modes in an optical interferometer. So, for example, for modes 1 and 2 we abbreviate $|x\rangle_{1}|y\rangle_{2}$ as $|x,y\rangle$. These states are of interest due to the fact that they greatly improve the resolution and sensitivity of interferometry for metrology. Also, it has been shown that they would improve the resolution with which lithographic features may be written. We show that (monochromatic) N00N state absorption fares poorly as $N$ increases. Thus when
considering possible applications of these states, such as to quantum
lithography \cite{Boto,lith} or metrology \cite{Dowling,met}, we need to keep
an eye towards maximizing these absorption rates by varying their spectral
parameters. We do this knowing that it has been found that squeezed light can exhibit novel two photon absorption properties, such as linear growth of absorption rate with intensity and decreasing absorption, for increasing field \cite{Gea, line, proof}.

We consider in detail the case of a $|2::0\rangle$ state used in a quantum lithography or quantum metrology setup. We include spectral information and derive a general expression for the two-photon absorption probability. Then we numerically maximize the probability function and find the setup which maximizes the two-photon absorption. The absorption probability can be improved by several orders of magnitude by carefully adjusting the filter bandwidths, pump pulse length, and the length of the crystal. We go beyond most previous studies in that we obtain the two photon absorption probability directly, instead of only considering the second-order correlation function. 

In Section II we compare the absorption properties of ideal N00N states to other states of light. In section III we calculate the biphoton amplitude in the general case and then examine the absorption properties of this type of light in two regimes: the pulse-pumped, and the continuous-wave-pumped. Some of the more lengthy calculations are left to the appendices.

\section{Absorption Properties of Ideal N00N States}
Initially one may consider the ideal N00N state: $|N::0\rangle$ \cite{Boto}. This state contains no spectral information. It is an abstraction which can only exist in an optical cavity. Nonetheless it will provide some insight into how the absorption properties of N00N states compare to other sources. 

Agarwal  studied how multiphoton absorption rates are influenced by the specific properties of the incident light \cite{Agarwal}. He found that the equation of motion for the field can be written as

\begin{eqnarray}
\frac{\partial\langle\hat{a}^{\dagger}\hat{a}\rangle}{\partial t}=-2n\lambda^{(n)}\langle\hat{a}^{\dagger n}\hat{a}^{n}\rangle .\nonumber
\end{eqnarray}

\noindent Here, $\lambda^{(n)}$ is the absorption coefficient for the $n$-photon absorption  process and contains information about the medium that is acting as the absorber. The rate of change of the number of photons in the field
is proportional to the absorption rate. So the probability of an
$n$-photon absorption event occurring is,

\begin{eqnarray}
P_{n}=\kappa\frac{\langle\hat{a}^{\dagger n}\hat{a}^{n}\rangle}{\langle\hat{a}^{\dagger}\hat{a}\rangle^{n}},\nonumber
\end{eqnarray}

\noindent where $\kappa$ is some constant which we set to one in the interest of simplicity. This represents roughly the probability of $n$ photons in a light field being absorbed by an $n$-photon resist. We can use this information to produce a graph to see how
the multi-photon absorption probability of N00N states scale with $N$, when
compared to quantum states that are not path entangled such as thermal, number, and coherent. 

\begin{figure}
\includegraphics{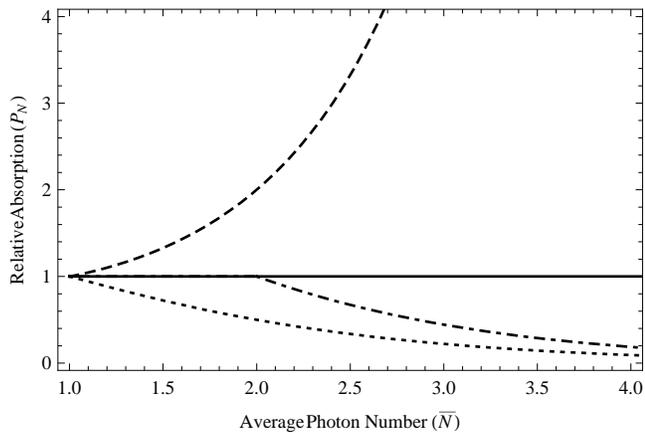}
\caption{\label{multi1} Multiphoton absorption probability of various
sources as a function of the average number of photons in the field
($N^{th}$ order absorption vs. $\bar{N}$ average photons). All probabilities
normalized relative to coherent states (solid line). The dashed line
is a thermal state. The dotted line is a Fock (number) state, and
the dot-dashed line is a N00N state. }
\end{figure}

Thermal states are described by the following density matrix

\begin{eqnarray}
\hat{\rho}_{\mathrm{thermal}}=\frac{1}{Z}\sum^{\infty}_{j=0}e^{-E_{j}/kT}|j\rangle\langle j|, \nonumber \\
Z=\frac{e^{-\hbar\omega /2kT}}{1-e^{-\hbar\omega /kT}},\quad E_{j}=\hbar\omega (j+\frac{1}{2}).\nonumber
\end{eqnarray}

\noindent Obtaining the matrices for number, coherent, and N00N states is straightforward. For a two mode state the annihilation operator is given as $\hat{a}=\hat{a}_{1}+\hat{a}_{2}$ where one and two label the two paths the photon may take \cite{Boto}.

See Fig. \ref{multi1}. Thermal states clearly have the greatest rates of multiphoton absorption. This can be attributed to the fact that thermal states
exhibit bunching. That is that photons from thermal radiation
tend to be tightly correlated in time. Fock (number) states fare the worst. This feature of Fock states is
connected to the fact that number states represent standing waves
where the locations of the individual photons are evenly spaced out (or
anti-bunched) in space and time with no definable phase.
The multiphoton absorption properties of coherent states stay
constant with respect to photon number. Since photons in coherent
states are randomly dispersed in space and time the chance of two or
more photons being correlated is simply proportional to the
intensity (average photon number). Since we are considering how
$n$-photon absorption scales against average photon number, the graph
is flat, providing a convenient measuring stick to gauge other
fields.

For N00N states the multiphoton absorption behaves as

\begin{eqnarray}
P_{N}&=& 1 \quad\quad \quad \mathrm{for} \quad N=1, \nonumber \\ 
&=&2\frac{N!}{N^{N}} \quad \mathrm{for} \quad N\geq 2,\nonumber
\end{eqnarray}

\noindent where $\kappa$ has been set to one. N00N states fare a factor of two better than Fock
states, although absorption rates are still far from optimal. The reason N00N sates have this factor of two is due to the path entanglement. Mathematically this two comes from the normalization constant that path entanglement requires.  

These results seem to reinforce the findings of Tsang in Ref. \cite{Tsang}: generally the absorption properties of N00N states are poor. It should be re-emphasized that these states, containing no spectral or temporal
information, are idealizations. Fig. \ref{multi1} can only be seen
as providing a rough idea of how absorption scales. 

Quantum lithography or metrology will only be useful if the detector or material needs to be exposed to the field for a reasonable period of time. The above ideal-field results seem to make this unlikely. However, they show only that the quantum mechanical properties (i.e. the bare state vector) of N00N states lead to poor absorption rates. They say nothing about how the spectral properties of realistic N00N state pulses effect the multiphoton absorption probability. We thus examine in detail a specific well known case: the $|2::0\rangle $ state. Though this state is not practical for metrology or lithography its optimization would provide a proof of principle that the absorption properties of higher $N$ states could be improved in a similar manner.  

\section{Absorption Properties of Realistic $|2::0\rangle $ States}

It is possible to write realistic states, which include the spectral information of the light of interest. Furthermore,
after these states are used to obtain absorption rates, the
arrangement of optical elements which optimizes absorption can be
found. 

Several works have examined how biphotons produced by parametric down conversion or electromagnetically induced transparency may be compressed or otherwise modified so that they exhibit tighter correlations \cite{Harris,Du,new,dayan1,Teich,torres}. There is an excellent paper by 
Dayan which studies the properties of a semi-stationary, undepleted beam of squeezed light produced by a spectrally narrow pump \cite{dayan2}.

\begin{figure}
\includegraphics[scale=0.145]{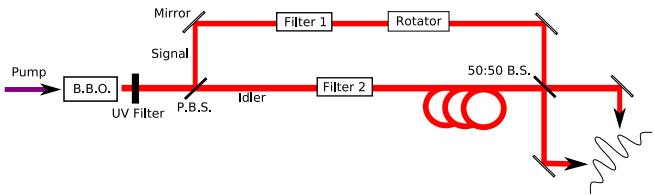}
\caption{\label{setup} The basic setup. A nonlinear crystal (BBO in this case) creates a degenerate pair of photons. Each photon is subjected to a filter. A polarization rotator ensures that the two photons are indistinguishable. A beam splitter creates a $|2::0\rangle $ state that results in an interference pattern. }
\end{figure}

We consider the case of a
$|2::0\rangle $ state used in a quantum
lithography \cite{Boto,lith} or remote quantum metrology \cite{Dowling,met}
setup. The $|2::0\rangle $ state we investigate is produced by co-linear type II degenerate down conversion and a beam splitter (see Fig. \ref{setup}). This setup is very simple, but by tuning these few basic optical elements we can see a large improvement in two photon absorption rates, without recourse to exotic techniques. We make no assumptions about stationarity or about the relative sizes of the field bandwidths (apart from one very broadly applicable assumption \--- that the field's bandwidth is narrower than the atomic transition frequencies; this will be discussed further in Appendix B).   

\subsection{The Type-II Biphoton}
The output state of the crystal during type-II down conversion is described by \cite{Rubin}

\begin{eqnarray}
|\Psi\rangle\lefteqn{=C\sum_{kk'}\int^{\infty}_{0} d\omega_{p}{\int_{0}}^{L}dz e^{-\mathcal{D}\left(\frac{\omega_{p}-\Omega{p}}{\sigma_{p}}\right)^{2}}e^{iz\Delta_{kk'}}}\nonumber \\
& &\times \delta(\omega_{ok}+\omega_{ek'}-\omega_{p}){\hat{a}_{ok}}^{\dagger}{\hat{b}_{ek'}}^{\dagger}|0\rangle_{o}|0\rangle_{e}\label{SPDC}
\end{eqnarray}

\noindent for a co-linear pump. The sum extends over all possible
wavevector modes. A polarizing beam splitter separates the ordinary and extraordinary beams into different spatial modes. The operators $\hat{a}$ and $\hat{b}$ represent these modes. $\Delta_{kk'}$ is the phase mismatch, defined:
$k_{p}-k-k'$, $\omega_{p}$ is the frequency of the pump laser, and
the $z$ integral extends over the length of the crystal.
$\Omega_{p}$ and $\sigma_{p}$ are the central frequency and the spectral FWHM
of the pump, respectively. $e$ and $o$ label the extraordinary and ordinary beams. The factor $\mathcal{D}$ is defined as $4\mathrm{ln}(2)$. We are assuming the pump laser is Fourier-transform-limited.

This state corresponds to squeezing just above threshold, such that mainly $|0,0\rangle$ and $|1,1\rangle$ are produced. The $|0,0\rangle
$ term is then dropped because it can not effect the two photon absorption process. 

Since the two beams are distinguishable after the polarizing beam splitter, one of the beams must be subject to a polarization rotator in order for it to be made indistinguishable from the other. The beams must be indistinguishable so that the $|1,1\rangle$ states interfere destructively and produce $|2::0\rangle $ \cite{HOM}. 

To study the two photon absorption probability of this state we utilize the well confirmed \cite{con1}\cite{con2} Eqs. (2.15) and (2.16) from Mollow \cite{Mollow}

\begin{eqnarray}
P_{2}\lefteqn{=\int\int d\omega ' d\omega g^{*}(\omega ')}\nonumber \\
& &\times S^{(2)}(\omega_{f}-\omega ',\omega ';\omega_{f}-\omega,\omega)g(\omega). \label{p2}
\end{eqnarray}

\noindent Where $g(\omega)$ is the atomic response function, $\omega_{f}$ is
the frequency of the final state, and $S^{(2)}$ is the spectral correlation function (the fourier transform of the temporal correlation function), which in our case is

\begin{eqnarray}
S^{(2)}(\omega_{1} ',\omega_{2} ';\omega_{1},\omega_{2})=\mathcal{Z}(\omega_{1}',\omega_{2}')^{*}\mathcal{Z}(\omega_{1},\omega_{2}),\label{S}
\end{eqnarray}

\noindent where,

\begin{eqnarray}
\mathcal{Z}(\omega_{1},\omega_{2})\equiv\int\int dt_{1}^{d}dt_{2}^{d}e^{i\omega_{1}t_{1}^{d}}e^{i\omega_{2}t_{2}^{d}}A(t_{1}^{d},t_{2}^{d})
.\nonumber 
\end{eqnarray}

\noindent Here, $\omega_{1}'$ and $\omega_{2}'$ represent the
negative frequency components of the field associated with
${t_{1}'}^{d}$ and ${t_{2}'}^{d}$. The factors $\omega_{1}$ and $\omega_{2}$
represent the positive frequency components of the field associated
with ${t_{1}}^{d}$ and ${t_{2}}^{d}$. The above equation does not include
the effects of natural linewidth, which will be discussed later.
We should note that other equations, also derived by Mollow
\cite{Mollow}, assume that the field is stationary, something which
is not true in general for our calculation here. $A$ is the biphoton amplitude, defined below.

We start with the two-photon correlation function \cite{Glauber}

\begin{eqnarray}
G^{(2)}=\langle\Psi|\hat{E}^{(-)}({t_{1}'}^{d})\hat{E}^{(-)}({t_{2}'}^{d})\hat{E}^{(+)}({t_{1}}^{d})\hat{E}^{(+)}({t_{2}}^{d})|\Psi\rangle .\label{G}
\end{eqnarray}

\noindent The primed and unprimed time variables represent the possibility of
the biphoton traveling via two different paths of different lengths,
as is the case in an interferometric setup, and $d$ labels the times
as being detection times. $\hat{E}^{(+)}(t)$ is the positive-frequency
electric field operator defined by

\begin{eqnarray}
\hat{E}^{(+)}({t_{j}}^{d})&=&i\sum_{s_{j}k_{j}}\left(\frac{\hbar\omega_{k_{j}}}{2\epsilon_{0}V}\right)^{1/2} e^{-\mathcal{D}\left(\frac{\omega_{k_{j}} - \Omega_{f}}{\sigma_{f}}\right)^{2}} \nonumber \\
& &\times e^{-i\omega_{k_{j}}{t_{j}}^{d}}\hat{a}_{s_{j}k_{j}}(0).\label{E}
\end{eqnarray}

\noindent Where the approximation $e^{\pm i \bold{k}\cdot \bold{r}}\approx 1$ has been made and $s$
denotes either horizontal or vertical polarization. The time $t^{d}=0$ is defined as the time the photon is created. Note that $\hat{E}^{(+)\dagger}=\hat{E}^{(-)}$. Also, $\Omega_{f}$, and $\sigma_{f}$ are the
central frequency and the FWHM of the filter in a specific arm, respectively.

Now, by inserting a complete set of number states in the correlation
function, and observing that all but the $|0\rangle\langle 0|$ term
will cancel, we can rewrite Eq. (\ref{G}) as

\begin{eqnarray}
G^{(2)}\lefteqn{={\langle \Psi |\hat{E}^{(-)}({t_{1}'}^{d})\hat{E}^{(-)}({t_{2}'}^{d})|0\rangle}}\nonumber \\
& &\times {\langle 0 |\hat{E}^{(+)}({t_{1}}^{d})\hat{E}^{(+)}({t_{2}}^{d})|\Psi\rangle}\nonumber \\
\lefteqn{\equiv {A({t_{1}'}^{d},{t_{2}'}^{d})}^{*}A({t_{1}}^{d},{t_{2}}^{d})}.\nonumber
\end{eqnarray}

\noindent The above equation defines the biphoton amplitude $A({t_{1}}^{d},{t_{2}}^{d})$. The expressions for $A$ for SPDC were first
calculated by Keller and Rubin in Ref. \cite{Keller} and elaborated upon in Refs.\cite{Kim,thesis,shih}. We follow their
calculations somewhat closely. For more details see Appendix A. The result is $A(t_{1}^{d},t_{2}^{d})=\mathcal{A}(t_{1}^{d},t_{2}^{d})+\mathcal{A}(t_{2}^{d},t_{1}^{d})$, where the script $\mathcal{A}$ represents the biphoton amplitude for down conversion followed by a polarizing beam splitter. The italic $A$ is the biphoton amplitude for our setup. The result of our calculation for $\mathcal{A}$ is

\begin{eqnarray}
\mathcal{A}({t_{1}}^{d},{t_{2}}^{d})&=&e^{-i(\Omega_{e}{t_{2}}^{d}+\Omega_{o}{t_{1}}^{d})}e^{-\frac{({t_{1}}^{d}O_{U}\sigma_{o}+{t_{2}}^{d}E_{U}\sigma_{e})^{2}}{4\mathcal{D}U^{2}}}\nonumber \\ 
& &\times\frac{U_{e}U_{o}U_{p}\sigma_{e}\sigma_{o}\sigma_{p}}{U\sqrt{\mathcal{D}}}\nonumber \\
& &\times\left[\mathrm{Erf}(\mathcal{T})-\mathrm{Erf}(\mathcal{T}+l)\right]. \label{biphoton}
\end{eqnarray}

\noindent Where

\begin{eqnarray}
\mathcal{T}&=&\frac{({t_{1}}^{d}-{t_{2}}^{d})P_{U}\sigma_{e}\sigma_{o}+{t_{1}}^{d} E_{U}\sigma_{o}\sigma_{p}-{t_{2}}^{d}O_{U}\sigma_{e}\sigma_{p}}{2U\sqrt{\mathcal{D}({\sigma_{e}}^{2}+{\sigma_{o}}^{2}+{\sigma_{p}}^{2}})}\nonumber \\
l&=&\frac{LU}{2U_{e}U_{o}U_{p}\sqrt{\mathcal{D}({\sigma_{e}}^{2}+{\sigma_{o}}^{2}+{\sigma_{p}}^{2}})}.\nonumber
\end{eqnarray}

\noindent And

\begin{eqnarray}
P_{U}=U_{p}(U_{e}-U_{o})\sigma_{e}\sigma_{o},\quad E_{U}=U_{e}(U_{p}-U_{o})\sigma_{p}\sigma_{o}\nonumber \\
O_{U}=U_{o}(U_{e}-U_{p})\sigma_{e}\sigma_{p}, \quad
U^{2}={P_{U}}^{2}+{E_{U}}^{2}+{O_{U}}^{2}. \nonumber
\end{eqnarray}

\noindent The error function is commonly defined as

\begin{eqnarray}
\mathrm{Erf}(x)=\frac{2}{\sqrt{\pi}}\int_{0}^{x}dy e^{-y^{2}}.\nonumber
\end{eqnarray}

\noindent The terms $U_{e}$, $U_{o}$, and $U_{p}$ represent the group velocities of the extraordinary, ordinary, and pump beams, respectively. The factors $\sigma_{e}$ and $\sigma_{o}$ are the bandwidths of the filters in the arms of the interferometer. The term $\sigma_{p}$ is the bandwidth of the pump. The factors $\Omega_{e}$ and $\Omega_{o}$ are the central frequencies of the extraordinary and ordinary beams. For now the normalization constant has been left off for the sake of simplicity as it will only effect the height of the amplitude, not its overall shape.  

Hence we give the general biphoton amplitude in its most general form, which is a new result.

In order to evaluate this expression we must find the group velocities of the pump, ordinary and extraordinary beams inside of the $\chi^{(2)}$ crystal. The index of refraction as a function of wavelength can be found using the Sellmeier equations. Below are the Sellmeier equations for $\beta$-Barium Borate (BBO) \cite{bbo}

\begin{eqnarray}
n_{o}(\lambda_{o})&=&\sqrt{2.7359+\frac{0.01878}{\lambda_{o}^{2}-0.01822}-0.0135\lambda_{o}^{2}}\nonumber \\
n_{e}(\lambda_{e})&=&\sqrt{2.3753+\frac{0.01224}{\lambda_{e}^{2}-0.01667}-0.01516\lambda_{e}^{2}}.\nonumber
\end{eqnarray}

\noindent Where the wavelength is given in micro-meters. In type-II down conversion in BBO the pump beam experiences the same index of refraction as the extraordinary beam. It is also important to consider the angle the beams form with respect to the optic axis of the crystal. Since we are investigating the degenerate co-linear case with planar phase matching the optic axis must be set to be $42.4^{\circ}$ off the pump beam's direction of propagation for a $400\mathrm{nm}$ pump (the ordinary and extraordinary beams are co-linear with the pump and selected with a pinhole downstream) \cite{thesis}. For the pump and extraordinary beams we must use the effective index of refraction, given by

\begin{eqnarray}
n^{\mathrm{eff}}(\lambda_{e},\phi)=\left[\frac{\cos^{2}(\phi)}{n_{o}^{2}(\lambda)}+\frac{\sin^{2}(\phi)}{n_{e}^{2}(\lambda)}\right]^{-\frac{1}{2}},\nonumber
\end{eqnarray}

\noindent where $\phi$ is the angle between the beam and the optic axis of the crystal. We can now calculate the group velocity

\begin{eqnarray}
U_{e,o,p}(\lambda_{e,o,p})=\left(\frac{n^{\mathrm{eff}}_{e,o,e}(\lambda_{e,o,p})}{c}-\frac{\lambda_{e,o,p}}{c}\frac{\partial n^{\mathrm{eff}}_{e,o,e}(\lambda_{e,o,p})}{\partial \lambda_{e,o,p}}\right)^{-1}.\nonumber
\end{eqnarray}

\noindent And $n_{o}^{\mathrm{eff}}=n_{o}$. We then find for the degenerate case
($\Omega_{o}=\Omega_{e}$) for $\lambda_{p}=400\mathrm{nm}$ that
$U_{o}(\Omega_{o})=1.781\times 10^{8}\mathrm{m/s}$,
$U_{p}(\Omega_{p})=1.756\times 10^{8}\mathrm{m/s}$ and $U_{e}(\Omega_{e})=1.845\times 10^{8}\mathrm{m/s}$.

\begin{figure}
\begin{picture}(250,250)
\put(88,0){\Large{$t_{1}^{d}$($10^{-13}\mathrm{sec}$)}}
\put(0,88){\begin{sideways}\Large{$t_{2}^{d}$($10^{-13}\mathrm{sec}$)}\end{sideways}}
\put(20,20){\includegraphics[scale=0.8]{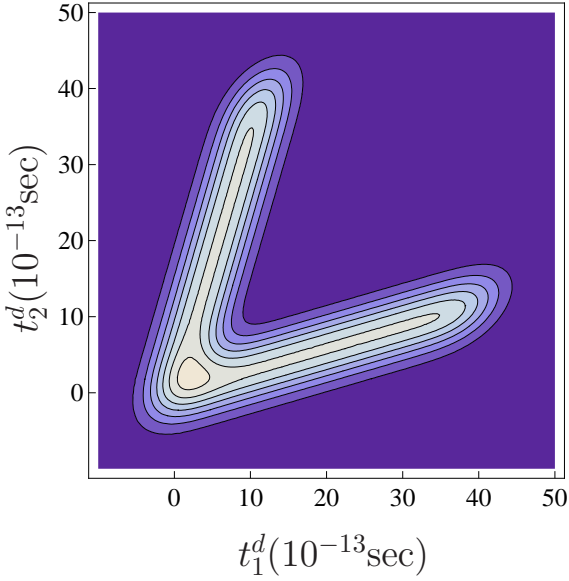}}
\end{picture}
\caption{Contour plot of the absolute value of the biphoton amplitude for our setup $|A|$. $\sigma_{e}=\sigma_{o}=\sigma_{p}=10^{13}\mathrm{Hz}$ and $L=1.5\mathrm{cm}$. $t_{1}^{d}$ and $t_{2}^{d}$ are in units of $10^{-13}\mathrm{sec}$ \label{density1}}
\end{figure}

Fig. \ref{density1} contains a contour plot of the absolute value of the biphoton amplitude for our setup ($A$). This graph displays an interesting splitting which is symmetric about a line given by $t^{d}_{1}-t^{d}_{2}=0$. Each point on this line represents a different average arrival time of the biphoton $(t^{d}_{1}+t^{d}_{2})/2$, a line drawn perpendicular to this line of symmetry represents another axis. That axis defines an entanglement time (the temporal distance between the two photons) as $t^{d}_{1}-t^{d}_{2}$. The symmetric splitting represents the fact that two photons generated far away from the exit-surface of the crystal will drift apart in time. Thus, as the average arrival time increases (a delay being indicative of more time spent in the crystal) the photons drift apart. The symmetry is a result of the interferometer scrambling the information corresponding to which photon took which path. So for a set average arrival time there is an equal probability that the $e$ photon will arrive first or that the $o$ one will.     

We can check Eq. (\ref{biphoton}) by taking limits and comparing to known formulae.
Taking the limit of the biphoton amplitude as $\sigma_{o}$ and $\sigma_{e}$ go to infinity (the case of no filtering) and keeping in mind that one of the definitions of the Heaviside step function is,

\begin{eqnarray}
H(x)=\lim_{k\rightarrow\infty}\frac{1}{2}\left(1+\mathrm{Erf}(kx)\right).\label{H}
\end{eqnarray}

\noindent We are left with

\begin{eqnarray}
\mathcal{A}({t_{1}}^{d},{t_{2}}^{d})&=&e^{-\sigma_{p}^2J}\left[2H\left(A-\frac{B}{2}\right)-2H\left(A+\frac{B}{2}\right)\right], \nonumber
\end{eqnarray}

\noindent where

\begin{eqnarray}
A&=&\frac{(t_{1}^{d}-t_{2}^{d})U_{e}U_{o}+\frac{1}{2}L(U_{e}-U_{o})}{2U_{e}U_{o}\sqrt{\mathcal{D}}}, \nonumber \\
B&=&\frac{L(U_{e}-U_{o})}{2U_{e}U_{o}\sqrt{\mathcal{D}}},\nonumber\\
J&=&\frac{\left[t_{1}^{d}U_{o}(U_{e}-U_{p})+t_{2}^{d}U_{e}(U_{p}-U_{o})\right]^{2}}{\sqrt{\mathcal{D}}(U_{e}-U_{o})^{2}U_{p}^{2}}.\nonumber
\end{eqnarray}

\noindent Utilizing the identity

\begin{eqnarray}
\mathrm{Rect}\left(\frac{x}{\tau}\right)=H\left(x+\frac{\tau}{2}\right)-H\left(x-\frac{\tau}{2}\right),\nonumber
\end{eqnarray}

\noindent where

\begin{eqnarray}
\mathrm{Rect}(x)=\left\lbrace \begin{array}{ccc}
0\quad\mathrm{if}\quad |x|>0.5, &\\
\frac{1}{2}\quad\mathrm{if}\quad |x|=0.5,&\\
1\quad\mathrm{if}\quad |x|<0.5.& \end{array}\right.\nonumber
\end{eqnarray}

\noindent we obtain the expression

\begin{eqnarray}
\mathcal{A}({t_{1}}^{d},{t_{2}}^{d})=-2e^{-\sigma_{p}^2J}\mathrm{Rect}\left(\frac{A}{B}\right).\nonumber
\end{eqnarray}

\noindent Which is equivalent to the expression given by Kim, et al., in Ref. \cite{thesis}.

Eq. (\ref{p2}) given together with Eq. (\ref{up}) gives all the necessary information for calculating the two-photon absorption probability for N00N states of $N=2$. Details of this calculation are given in Appendix B. The result is

\begin{eqnarray}
\widetilde{P}_{2}&=&\frac{C''}{L\sigma_{p}}\int^{\infty}_{-\infty}d\nu_{f}e^{-2\mathcal{D}{\nu_{f}}^{2}\left(\frac{(u_{e}+U_{2})^{2}}{\sigma_{e}^{2}U_{2}^{2}}+\frac{u_{e}^{2}}{\sigma_{o}^{2}U_{2}^{2}}+\frac{1}{\sigma_{p}^{2}}\right)}\nonumber\\
& &\times \frac{\left| \mathrm{Erf}\left(\mathcal{E}\nu_{f}\right) -\mathrm{Erf}\left(\mathcal{E}\nu_{f}-\mathcal{L}\right) \right|^{2}}{\left[1+4\left(\frac{\nu_{f}}{\kappa_{f}}\right)^{2}\right].\label{p2t}}
\end{eqnarray}

\noindent Where

\begin{eqnarray}
u_{e}&=&\frac{1}{U_{p}}-\frac{1}{U_{e}},\nonumber\\
U_{2}&=&\frac{1}{U_{o}}-\frac{1}{U_{e}},\nonumber\\
\mathcal{E}&=&\frac{i\sqrt{\mathcal{D}}(U_{2}\sigma_{o}^{2}+u_{e}(\sigma_{e}^{2}+\sigma_{o}^{2}))}{U_{2}\sigma_{e}\sigma_{o}\sqrt{\sigma_{e}^{2}+\sigma_{o}^{2}}} ,\nonumber \\
\mathcal{L}&=&\frac{LU_{2}\sigma_{e}\sigma_{o}}{2\sqrt{\mathcal{D}(\sigma_{e}^{2}+\sigma_{o}^{2})}},\nonumber\\
C''&=&\frac{8}{U_{2}}(1+\sqrt{\pi})^{2}\sqrt{2\pi^{7}\mathcal{D}}\label{el}
\end{eqnarray}

\noindent And $\kappa_{f}$ is the FWHM of the final state. The atomic response function $g(\omega)$ can be moved outside the integral for the light sources we consider (see Appendix B for details). Unfortunately this integral is intractable analytically, we therefore perform the integration over $\nu_{f}$ numerically.

\subsection{Numerical Calculation of Type-II Absorption Rate}
We wish to use Eqs. (\ref{p2},\ref{p2t},\ref{el}) to calculate the relative absorption
rate for type-II down conversion. After this is done, we can use the
information to maximize absorption by adjusting the available
spectral parameters. 

\subsubsection{Pulse-Pumped Case}

We start with the case of a pulsed pump. Since the two-photon absorption probability is a complicated quantity with five adjustable parameters ($\sigma_{e}$, $\sigma_{o}$, $\sigma_{p}$, $L$, and $\kappa_{f}$), we present several graphs to help elucidate the structure of this mathematical object. All of the graphs are scaled so that the maximal absorption probability \--- within the parameters we consider \--- is set to one. Thus $\widetilde{P}_{2}=1$ does not represent an absorption probability of unity. The normalization is consistent across all the graphs so that they may be compared to one another on the same scale.   

\begin{figure}
\begin{picture}(250,180)
\put(0,70){\Large{$\widetilde{P}_{2}$}}
\put(205,40){\large{$\mathrm{Log}(\sigma_{e})$}}
\put(55,16){\large{$\mathrm{Log}(\sigma_{o})$}}
\put(10,0){\includegraphics[scale=0.7]{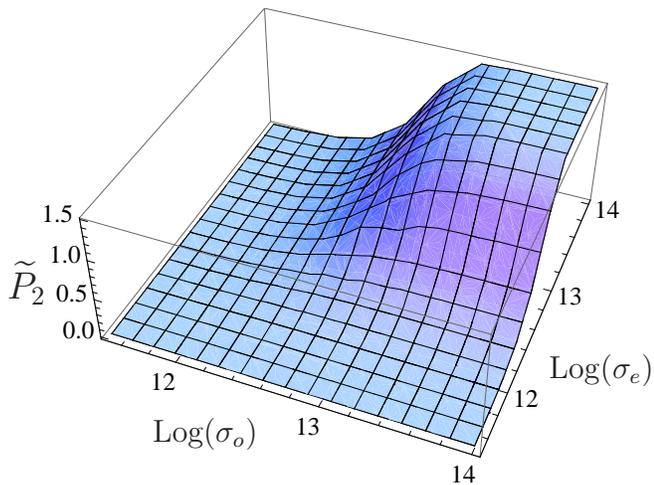}}
\end{picture}
\caption{\label{seo}A plot of the scaled two-photon absorption probability for the realistic $|2::0\rangle $ state as a function of the logs (base 10) of the bandwidths of the filters in the arms of the interferometer (in Hz). $L=2.3\mathrm{mm}$, $\sigma_{p}=10^{12}\mathrm{Hz}$, and $\kappa_{f}=10^{14}\mathrm{Hz}$.}
\end{figure}

Fig. \ref{seo} is a plot of the two-photon absorption as a function of the base-ten logarithms of the filter bandwidths. We see that a wider filter is preferable. This is because if the spectrum of the light is wide then the temporal distribution will be narrow, increasing the probability that the photons will arrive close together, triggering a two-photon absorption event.     

\begin{figure}
\begin{picture}(250,160)
\put(116,3){\Large{$\mathrm{Log}(\sigma_{p})$}}
\put(0,90){\Large{$\widetilde{P}_{2}$}}
\put(52,71){$\kappa_{f}=10^{11}\mathrm{Hz}$}
\put(113,90){$10^{12}$}
\put(161,118){$10^{14}$}
\put(20,20){\includegraphics[scale=0.91]{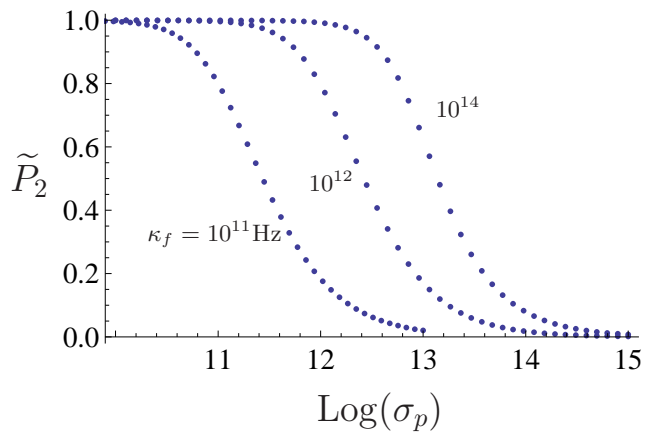}}
\end{picture}
\caption{\label{sp}A plot of the scaled two photon absorption probability for the realistic $|2::0\rangle$ state as a function of the log (base 10) of the bandwidth of the pump (in Hz) for three separate settings of the width of the final state of the absorber (in Hz). $L=2.3\mathrm{mm}$ and $\sigma_{e}=\sigma_{o}=10^{13}\mathrm{Hz}$.}
\end{figure}

Fig. \ref{sp} is a plot of the two-photon absorption probability as a function of the logarithm base ten of the bandwidth of the pump for several different atomic linewidths. Clearly a broader atomic linewidth will result in better absorption, as this will allow a greater range of frequency pairs to be absorbed. For the pump beam a smaller bandwidth is preferable. Once the bandwidth of the pump approaches that of the filters, or the linewidth of the atom, the absorption drop off is dramatic. Furthermore a narrower pump bandwidth constrains the frequencies of the daughter photons, increasing the probability that the sum of their energies will be resonant.   

\begin{figure}
\begin{picture}(250,160)
\put(117,3){\Large{$L(\mathrm{mm})$}}
\put(2,90){\Large{$\widetilde{P}_{2}$}}
\put(90,123){$\sigma_{e}=\sigma_{o}=10^{13}\mathrm{Hz}$}
\put(60,50){$4\times 10^{12}$}
\put(20,20){\includegraphics[scale=0.91]{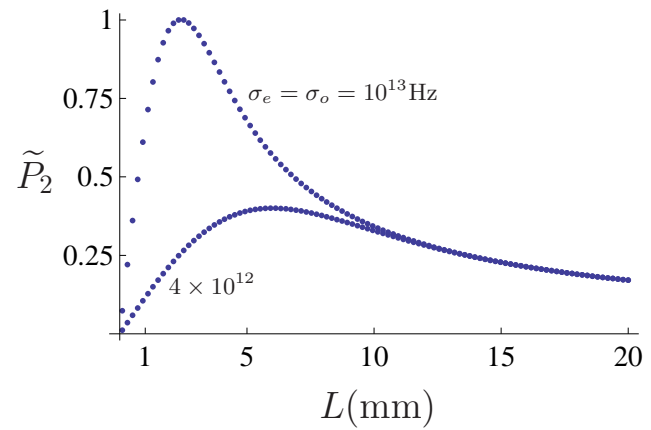}}
\end{picture}
\caption{\label{L1}A plot of the scaled two photon absorption probability for the realistic $|2::0\rangle$ state as a function of the length of the crystal for two different settings of the filters (in Hz). $\sigma_{p}=10^{12}\mathrm{Hz}$ and $\kappa_{f}=10^{14}\mathrm{Hz}$ }
\end{figure}

Fig. \ref{L1} is a plot of the two-photon absorption probability of a $|2::0\rangle$ pulse as a function of crystal length for two separate cases. This is perhaps the most interesting of the representations of Eq. (\ref{p2t}). The plot indicates the feature that for each setting of $\sigma_{o}$, $\sigma_{e}$, $\sigma_{p}$, and $\kappa_{f}$ there is an optimal crystal length that maximizes the probability of two-photon absorption. 

As light travels through the crystal the dispersive nature of the medium causes the two photons to drift apart in time. This effect results in a drop off in absorption as length increases. Conversely, if the crystal length is very short, the spectrum of the light will be very broad, and the filters will strongly limit the amount of light that can reach the absorber. For broad filters, any deviation from the optimal length causes a dramatic drop off in absorption. Information of this type would be useful when we design a two-photon quantum lithography experiment. A crystal cut to a specific length, for a given setup, would have the potential to enhance the two-photon absorption properties of the generated light. 

For example, take the case we have defined to be $\widetilde{P}_{2}=1$: Here we have chosen, $\sigma_{e}=\sigma_{o}=10^{13}\mathrm{Hz}$, $\sigma_{p}=10^{9}\mathrm{Hz}$, $L=2.3$ mm, and $\kappa_{f}=10^{14}\mathrm{Hz}$. For the same absorber ($\kappa_{f}$ the same), $\sigma_{e}=\sigma_{o}=10^{11}\mathrm{Hz}$, $\sigma_{p}=10^{13}\mathrm{Hz}$, and  $L=2$ cm, the absorption probability is is $\widetilde{P}_{2}=2.07\times 10^{-5}$.         

We may obtain completely analytical results for the case where there are no filters in the arms of the interferometer. Using Eq. (\ref{H}), Eq. ({\ref{p2t}) becomes

\begin{eqnarray}
\lim_{\sigma_{e},\sigma_{o}\rightarrow\infty}\widetilde{P}_{2}=\frac{C''\kappa_{f} e^{\frac{\mathcal{D}\kappa_{f}^{2}}{8\sigma_{p}^{2}}}}{2L\sigma_{p}}K_{0}\left(\frac{\mathcal{D}\kappa_{f}^{2}}{8\sigma_{p}^{2}}\right),\label{limit}
\end{eqnarray}

\noindent so long as $L\neq 0$. The function $K_{n}$ is the modified Bessel function of the second kind. Features of this function include a simple dependence on $L$ and a modified $\sigma_{p}$ dependence, as shown in Fig. \ref{limit}. Interestingly, for absorbers with very wide final states, the two photon absorption probability becomes almost independent of the bandwidth of the pump.  

\begin{figure}
\begin{picture}(250,160)
\put(117,3){\Large{$\sigma_{p}(\mathrm{Hz})$}}
\put(2,90){\Large{$\widetilde{P}_{2}$}}
\put(65,58){$\kappa_{f}=10^{11}\mathrm{Hz}$}
\put(138,85){$10^{12}$}
\put(190,145){$10^{13}$}
\put(20,20){\includegraphics[scale=0.7]{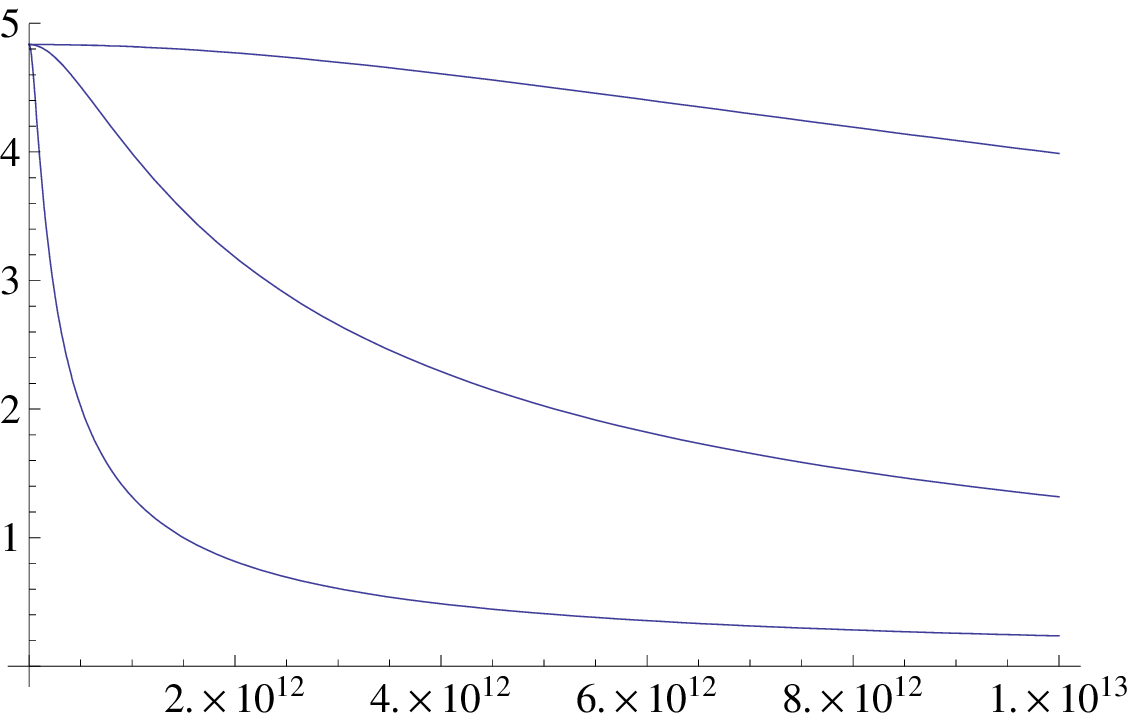}}
\end{picture}
\caption{\label{limit}A plot of the scaled two-photon absorption probability for the $|2::0\rangle$ state (in the limit of no filtering) as a function of the bandwidth of the pump. We indicate three separate settings of the width of the absorber. We also set $L=1$ mm. The scaling is the same as on the other pulse pumped plots.} 
\end{figure}

In all cases, given a particular $\kappa_{f}$, a balance must be struck between the related quantities of temporal and spectral correlation. Tight temporal correlation increases the probability of absorption. However this necessitates broad spectral distributions which have a lower probability of matching the correct energy for transition.

\subsubsection{Comparison to Coherent Light}

We can now tell how varying the spectral properties of $|2::0\rangle $ states will effect the absorption probability. However we have discovered nothing about how $|2::0\rangle $ light's absorption properties compare to other, more familiar, states of light. We would like to say wether $|2::0\rangle $ has any benefits or disadvantages when compared to, for example, coherent light. Coherent light has the advantage of being relatively well understood (or at least very extensively studied). Thus we shall attempt to find a state of coherent light which will serve as a fair comparison to $|2::0\rangle $, as produced in the fashion presented in this paper. We would like this fair state to have the same spectral profile as $|2::0\rangle $, but to lack the unique properties that the N00N state possesses: momentum entanglement and a high degree of temporal correlation. For details on the derivation of this state see Appendix C. The result for the two photon absorption probability is

\begin{eqnarray}
\widetilde{P}_{2}^{\alpha}&=&\pi I^{2}\mathrm{Erf}\left(\frac{Lu_{o}\sigma_{p}}{\sqrt{2\mathcal{D}}}\right)^{-2}\int^{\infty}_{-\infty}d\nu_{f}e^{-\frac{\mathcal{D}(u_{o}-u_{e})^{2}\nu_{f}^{2}}{u_{o}^{2}\sigma_{p}^{2}}}\nonumber\\
& &\times\frac{|\mathrm{Erf}(\mathcal{E}_{\alpha}\nu_{f})-\mathrm{Erf}(\mathcal{E}_{\alpha}\nu_{f}+\mathcal{L}_{\alpha})|^{2}}{\left[1+4\left(\frac{\nu_{f}}{\kappa_{f}}\right)^{2}\right]} .\label{p2coh}
\end{eqnarray}

\noindent Where

\begin{eqnarray}
\mathcal{E}_{\alpha}&=&\frac{i\sqrt{\mathcal{D}}(u_{o}-u_{e})}{\sqrt{2}u_{o}\sigma_{p}},\nonumber\\
\mathcal{L}_{\alpha}&=&\frac{Lu_{o}\sigma_{p}}{\sqrt{2\mathcal{D}}}.\nonumber\\
\end{eqnarray} 

\noindent and $u_{e}=\frac{1}{U_{p}}-\frac{1}{U_{e}}$ and $u_{o}=\frac{1}{U_{p}}-\frac{1}{U_{o}}$. Here, $I$ is the intensity of the light in each arm of the interferometer (i.e. $2I$ would be the total amount of light in the device).

This fair-comparison light can be though of as a very spectrally broad coherent source incident on a magic ``filter", which imposes on it the same spectral profile as the $|2::0\rangle $ state. Two spatial modes of this light are then used as the input into an interferometer without filters in the arms. The two-photon absorber is placed at the far end. 

The error functions in the the above coherent-state expression diverge much more rapidly than for the $|2::0\rangle$ case. Thus it is only possible to compute ${P}_{2}^{\alpha}$ for relatively narrow arguments of the integral over $\nu_{f}$, so we must consider a non-maximal setup as a test case. So for $\sigma_{p}=10^{9}\mathrm{Hz}$, $\kappa_{f}=10^{10}\mathrm{Hz}$, and $L=1$ cm we have    

\begin{eqnarray}
\widetilde{P}_{2}^{\alpha}=5.65\times 10^{-6}\widetilde{P}_{2}I^{2},\nonumber
\end{eqnarray}

\noindent where $\widetilde{P}_{2}$ has been calculated using Eq.  (\ref{limit}). 

If the intesities of the two kinds of light are set to be equal ($I=1$ in the above equation) then $|2::0\rangle $ is absorbed with a much higher probability than coherent light. This is directly a result of the frequency entanglement. Maximal absorption probability occurs when the photon energies add, such that the total is the same as the center of the final level. For SPDC it is insured that this condition will be closely met (exactly met in the case of a cw pump and a very thin crystal). However for the coherent analogue there is no such requirement, in fact the probability of two randomly chosen photons from the coherent pulses adding up to the resonance energy is miniscule. Even so, the above equation may be overly \textit{optimistic} about the absorption properties of coherent light. Recall that we were forced to consider narrow-band fields to make the calculation numerically tractable. The effect that causes coherent light to have a poor two-photon absorption probability will be worsened when the bandwidth is broader; a regime in which the $|2::0\rangle$ states' absorption probability improves. 

These positive results for the multiphoton absorption of entangled light is in seeming contradiction to the results of section II, which shows that coherent states should fair better than $|N::0\rangle $ states, where $N$-photon absorption is concerned. However that treatment does not take into account the spectral properties of highly quantum-mechanical states of light, which are likely to have high degrees of temporal correlation. 

In a paper by Tsang \cite{Tsang}, it was shown that the \textit{spatial} properties of N00N states cause them to have poor N-photon absorption rates. The relationship found in that paper is that N00N states have absorption rates that are lower than a classical analogue (in his case a monochromatic Fock state incident on a beam splitter) by a factor of $1/2^{N-1}$, due to spatial considerations independent of temporal correlations. He leaves open the question of whether time-domain effects can compensate for this effect. Given our above results it seems likely that the answer to this question is yes.      

However, the advantage entangled light enjoys is offset by the fact that SPDC has an extremely small intensity. Even relatively high production efficiencies are only approximately one pair per $10^{12}$ incident photons (see for example Ref. \cite{eff}). Furthermore it has been shown that correlated two photon absorption (absorption from two photons of the same pair as opposed to absorption from two photons of different pairs) from entangled light dominates only when the intensity is small \cite{trans}. This makes it even more important to improve absorption rates.     
  
\subsubsection{Continuous-Wave-Pumped Case}

Let us now investigate the case of a continuous-wave pump. The expressions derived in the previous section only make sense when considering a biphoton \textit{pulse}. They give the absorption probability of a biphoton of finite extent being absorbed as it passes an atom or other absorber. Note that since the pulses are Gaussian-like they are not strictly finite, but they are so closely temporally correlated that we may treat them as such. However the continuous wave case is stationary. We must now speak of a two photon absorption \textit{rate}.

First we recalculate the biphoton amplitude. Take Eq. (\ref{bi}), for a cw pump we set $\sigma_{p}=0$ and $\omega_{k_{o}}+\omega_{k_{e}}=\Omega_{p}$. We can then rewrite this equation as

\begin{eqnarray}
\mathcal{A}({t_{1}}^{d},{t_{2}}^{d})_{\mathrm{cw}}\lefteqn{=\sum_{k_{o}k_{e}}e^{-\mathcal{D}\left(\frac{\omega_{k_{o}}-\Omega_{o}}{\sigma_{o}}\right)^2}e^{-i\omega_{k_{o}}{t_{1}}^{d}}}\nonumber \\
& &\times e^{-\mathcal{D}\left(\frac{\omega_{k_{e}}-\Omega_{e}}{\sigma_{e}}\right)^2}e^{-i\omega_{k_{e}}{t_{2}}^{d}}\nonumber \\
& &\times \int^{L}_{0}dze^{iz\Delta_{k_{o}k_{e}}}\delta(\omega_{k_{o}}+\omega_{k_{e}}-\Omega_{p})\nonumber \\
\lefteqn{=\frac{e^{-i(\Omega_{o}{t_{1}}^{d}+\Omega_{e}{t_{2}}^{d})}}{U_{e}U_{o}}\int^{\infty}_{-\infty}d\nu\int^{L}_{0}dz}\nonumber \\
& &\times e^{-\mathcal{D}\left(\frac{\nu}{\sigma}\right)^{2}}e^{-i\nu({t_{2}}^{d}-{t_{1}}^{d})}e^{iz\Delta_{k_{o}k_{e}}},\nonumber
\end{eqnarray}   

\noindent where

\begin{eqnarray}
\sigma=\frac{\sigma_{e}\sigma_{o}}{\sqrt{\sigma_{e}^{2}+\sigma_{o}^{2}}}. \nonumber
\end{eqnarray}

\noindent In the second equality we have taken the continuous limit, changed variables from momentum to frequency, performed one of these integrals to eliminate the delta function, and then changed variables again to $|\nu_{e}|=|\nu_{o}|\equiv \nu $. We are able to do this due to conservation of energy for a cw pump: once the momentum of one photon is chosen the momentum of the other one is determined.  Also the phase mismatch becomes

\begin{eqnarray}
\Delta_{k_{e}k_{o}}=\nu\left(\frac{U_{e}-U_{o}}{U_{e}U_{o}}\right)\equiv u\nu . \nonumber
\end{eqnarray}

\noindent We now write in analogy to Eq. (\ref{up})

\begin{eqnarray}
\mathcal{A}({t_{1}}^{d},{t_{2}}^{d})_{\mathrm{cw}}&=&\frac{e^{-i(\Omega_{o}{t_{1}}^{d}+\Omega_{e}{t_{2}}^{d})}}{U_{e}U_{o}}\nonumber\\
& &\times\int^{L}_{0}dz\hat{\mathcal{F}}(\nu\rightarrow [{t_{2}}^{d}-{t_{1}}^{d}])Y(\nu) \nonumber \\
Y(\nu)&=&e^{-\mathcal{D}\left(\frac{\nu}{\sigma}\right)^2}e^{izu\nu}. \nonumber
\end{eqnarray} 

\noindent The solution for $A$ is easily found

\begin{eqnarray}
\mathcal{A}({t_{1}}^{d},{t_{2}}^{d})_{\mathrm{cw}}&=&\frac{e^{-i(\Omega_{o}{t_{1}}^{d}+\Omega_{e}{t_{2}}^{d})}}{U_{e}-U_{o}}\left[\mathrm{Erf}\left(\sigma\frac{({t_{1}}^{d}-{t_{2}}^{d})}{2\sqrt{\mathcal{D}}}\right)\right.\nonumber \\
& &-\left.\mathrm{Erf}\left(\sigma\frac{({t_{1}}^{d}-{t_{2}}^{d})-Lu}{2\sqrt{\mathcal{D}}}\right)\right].\nonumber
\end{eqnarray}

The absorption rate for stationary states is given by Eq. (3.17a) in Ref. \cite{Mollow} 

\begin{eqnarray}
w_{2}=2\left|g\left(\frac{1}{2}\Omega_{p}\right)\right|^{2}\int^{\infty}_{-\infty}dte^{2i\Omega_{p}t-\kappa_{f}|t|}G^{(2)}(-t,-t;t,t),\nonumber
\end{eqnarray}

\noindent where $\kappa_{f}$ is the width of the final state of the absorber. We have for the second-order correlation function,

\begin{eqnarray}
G^{(2)}(-t,-t;t,t)=\frac{2e^{-2i\Omega_{p}t}}{(U_{e}-U_{o})^{2}}\left|\mathrm{Erf}\left(\frac{Lu\sigma}{2\sqrt{\mathcal{D}}}\right)\right|^{2},\nonumber
\end{eqnarray}

\noindent and thus

\begin{eqnarray}
w_{2}=\frac{2\left|g\left(\frac{1}{2}\Omega_{p}\right)\right|^{2}}{(U_{e}-U_{o})^{2}}\left|\mathrm{Erf}\left(\frac{Lu\sigma}{2\sqrt{\mathcal{D}}}\right)\right|^{2}\int^{\infty}_{-\infty}dte^{-\kappa_{f}|t|}.\nonumber
\end{eqnarray}

\noindent The modulus squared of the error function is the only part that is dependent on the spectral properties of the field, so we fold the overall constants into the atomic response function, which we will then ignore. We then obtain the relatively simple result

\begin{eqnarray}
\widetilde{w}_{2}=\frac{1}{L}\left|\mathrm{Erf}\left(\frac{Lu\sigma}{2\sqrt{\mathcal{D}}}\right)\right|^{2},\label{cw}
\end{eqnarray} 

\noindent where the factor of $1/L$ comes from the state normalization. Unlike for the general case, the cw expression is fully analytical. Figs. \ref{cw2} and \ref{cw1} are plots of the absorption rate as a function of crystal length and filter bandwidths. The graphs have been normalized such that the greatest absorption rate for the range of parameters we consider is set to $\widetilde{w}_{2}=1$.

\begin{figure}
\begin{picture}(250,160)
\put(117,3){\Large{$L(\mathrm{mm})$}}
\put(2,90){\Large{$\widetilde{w}_{2}$}}
\put(90,143){$\sigma_{e}=\sigma_{o}=10^{13}\mathrm{Hz}$}
\put(57,123){$7.5\times 10^{12}$}
\put(70,93){$5\times 10^{12}$}
\put(90,60){$2.5\times 10^{12}$}
\put(20,20){\includegraphics[scale=0.87]{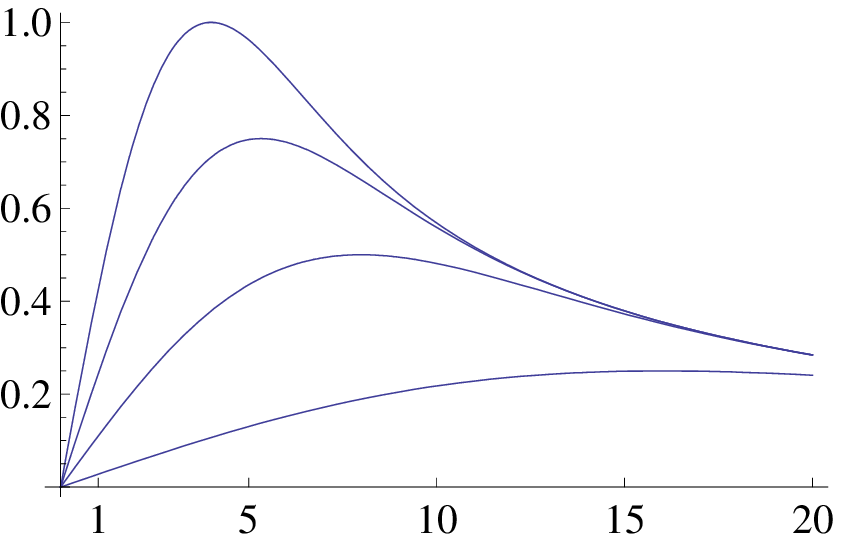}}
\end{picture}
\caption{\label{cw2} Plot of the two-photon absorption rate from a cw pumped crystal, as a function of crystal length, for four separate settings of the filters (in Hz).}
\end{figure}

\begin{figure}
\begin{picture}(250,190)
\put(2,97){\Large{$\widetilde{w}_{2}$}}
\put(27,20){\Large{$\sigma_{e}(10^{13}\mathrm{Hz})$}}
\put(180,25){\Large{$\sigma_{o}(10^{13}\mathrm{Hz})$}}
\put(20,10){\includegraphics[scale=0.78]{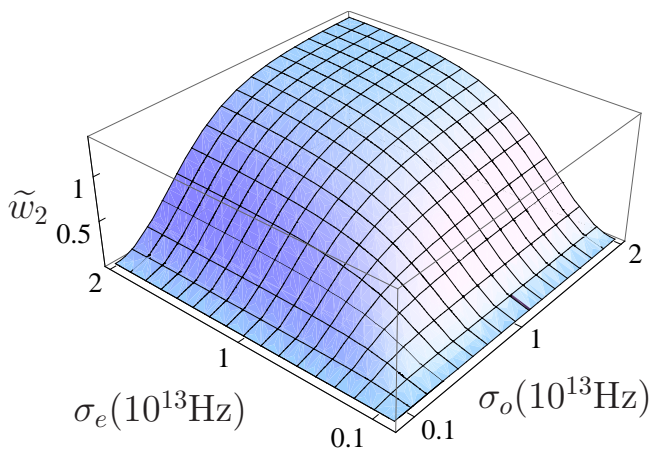}}
\end{picture}
\caption{\label{cw1} Plot of the two-photon absorption rate from a cw-pumped crystal, as a function of the bandwidths of the filters in the ordinary and extraordinary arms. Here $L=6$ mm.}
\end{figure}

So for a cw pump we have generally the same results as in the pulse-pumped case, with absorption improving as bandwidth increases. We also again observe that the graph of $\widetilde{w}_{2}$ as a function of $L$ displays peak values.

\smallskip

\section{Conclusion}
First, we analyzed the multiphoton absorption probability of states of the form $|N::0\rangle$ in the ideal case (with no spectral information). We found that the absorption probability scales poorly with $N$ when compared to coherent states, but well when compared to Fock states.

Second, we considered the case of a realistic $|2::0\rangle$ state produced by type-II spontaneous parametric down conversion, a polarization rotator and a beam splitter. We found that generally the two photon absorption properties of this states are highly dependent on the specific optical setup used. Using numerical methods it is possible to analyze the absorption probability as a function of the realistic parameters of the experiment. We can then adjust the optical setup of our model, including the length of the crystal, the bandwidths of the filters in the extraordinary and ordinary beams, and the pulse length of the pump. Running a maximization procedure over these variables it is possible to find the setup, which optimizes two-photon absorption. The difference between the optimal setup and a slight deviation may be dramatic.

This research constitutes a proof of principle of the idea that the poor production rates and detrimental spatial effects of highly quantum mechanical states of light may be mitigated by improving the absorption rates through spectral means. Though we consider in detail only the $|2::0\rangle$ case, it is likely that once methods of developing N00N states of higher $N$ are developed, similar methods may be applied to improve their $N$ photon absorption properties as well. 

Also, in a broader sense, similar techniques may be applied to increase the multiphoton absorption properties of any desirable state of light. All that is required is the quantum mechanical state vector. 

\section{Acknowledgements}
We would like to acknowledge the Army Research Office, the Boeing Corporation, The Foundational Questions Institute, the Intelligence Advanced Research Projects Activity, and the Northrup-Grumman Corporation for support. W.N.P. would also like to acknowledge the Louisiana Board of Regents for funding. Discussions with S. Vinjanampathy, S. Thanvanthri and G.S. Agarwal were invaluable.

\section*{Appendix A: Calculation of the Biphoton Amplitude}

We start with a field at the detector, given by Eq. (\ref{E}). Using the standard 50:50 beam splitter transformation we write out the electric field operator at the crystal

\begin{eqnarray}
E^{(+)}(t^{d})&=&\xi\sum_{k_{1}}\left(e^{-\mathcal{D}\frac{\omega_{k_{1}}^{2}}{\sigma_{o}^{2}}}\hat{a}_{k_{1}}+ie^{-\mathcal{D}\frac{\omega_{k_{1}}^{2}}{\sigma_{e}^{2}}}\hat{b}_{k_{1}}\right)\frac{e^{-i\omega_{1}\tau}}{\sqrt{2}}\nonumber \\
& &+\xi\sum_{k_{2}}\left(ie^{-\mathcal{D}\frac{\omega_{k_{2}}^{2}}{\sigma_{o}^{2}}}\hat{a}_{k_{2}}+e^{-\mathcal{D}\frac{\omega_{k_{2}}^{2}}{\sigma_{e}^{2}}}\hat{b}_{k_{2}}\right)\frac{e^{-i\omega_{2}\tau}}{\sqrt{2}}\nonumber
\end{eqnarray}

\noindent Where constants have been subsumed into the overall factor of $\xi$. The
polynomials $\omega_{k_{j}}$ have been ignored because they vary slowly when compared
to the exponential terms. The operators $\hat{a}$ and $\hat{b}$ are the annihilation operators acting on the first and second modes. The two mode annihilation operators are used to signify that the amplitude is dependent on the fields in both spatial modes. $\tau=t^{d}-l/c$ where $l$ is the distance between the crystal and the detector. We only consider the case where the interferometer is path balanced. This implies we
 calculate the absorption properties of
the central fringe. We assume that the absorption properties of the
other fringes will behave similarly. 

Now using the above and Eqs. (\ref{SPDC}, \ref{G}) we can find an expression for the biphoton amplitude of our setup

\begin{eqnarray}
A({t_{1}}^{d},{t_{2}}^{d})=\mathcal{A}({t_{1}}^{d},{t_{2}}^{d})+\mathcal{A}({t_{2}}^{d},{t_{1}}^{d})\label{bi}
\end{eqnarray}

\noindent where $\mathcal{A}$ is the biphoton amplitude for just the output of BBO and filters

\begin{eqnarray}
\mathcal{A}({t_{1}}^{d},{t_{2}}^{d})\lefteqn{=iC\xi^{2}\sum_{k_{o}k_{e}}e^{-\mathcal{D}\left(\frac{\omega_{k_{o}} - \Omega_{o}}{\sigma_{o}}\right)^{2}} e^{-i\omega_{k_{o}}{t_{1}}^{d}}}\nonumber \\
& &\times e^{-\mathcal{D}\left(\frac{\omega_{k_{e}} - \Omega_{e}}{\sigma_{e}}\right)^{2}}e^{-i\omega_{k_{e}}{t_{2}}^{d}}\nonumber \\
& &\times {\int_{0}}^{L}dz e^{-\mathcal{D}\left(\frac{\omega_{k_{o}}+\omega_{k_{e}}-\Omega{p}}{\sigma_{p}}\right)^{2}}e^{iz\Delta_{k_{o}k_{e}}}.\nonumber
\end{eqnarray}     

\noindent We have dropped the factor of $e^{-il/c}$ as it just introduces an overall phase. The central frequencies of the filters have been chosen to be the same as for the $e$ and $o$ rays. 

Now, define the following variables

\begin{eqnarray}
\nu_{o}\equiv \omega_{o}-\Omega_{o}, \nonumber\\
\nu_{e}\equiv \omega_{e}-\Omega_{e}, \nonumber\\
\nu_{p}\equiv \omega_{p}-\Omega_{p}. \label{coord}
\end{eqnarray}

\noindent Due to the delta function in Eq. (\ref{SPDC}), $ \Omega_{o}+\Omega_{e}=\Omega_{p}$ and $\nu_{o}+\nu_{e}=\nu_{p}$. Now, Taylor series expand the wavevector out to second order

\begin{eqnarray}
k(\nu_{j})&=&k(\Omega_{j})+\nu_{j}\left[\frac{dk(\nu_{j})}{d\omega_{j}}\right]_{\omega_{j}=\Omega_{j}}\nonumber \\
&=&k(\Omega_{j})+\frac{\nu_{j}}{U_{j}(\Omega_{j})}.\nonumber
\end{eqnarray}

\noindent Where $j=e,o,p$ and $U$ is the group velocity and $U=d\omega/dk$.
Taking into account that $k_{p}(\Omega_{p})=k_{e}(\Omega_{e})+k_{o}(\Omega_{o})$ The phase mismatch can now be rewritten as

\begin{eqnarray}
\Delta_{k_{o}k_{e}}&=&\frac{\nu_{p}}{U_{p}(\Omega_{p})}-\frac{\nu_{o}}{U_{o}(\Omega_{o})}-\frac{\nu_{e}}{U_{e}(\Omega_{e})} \nonumber \\ 
&=&\frac{\nu_{o}+\nu_{e}}{U_{p}(\Omega_{p})}-\frac{\nu_{o}}{U_{o}(\Omega_{o})}-\frac{\nu_{e}}{U_{e}(\Omega_{e})}.\label{mismatch}
\end{eqnarray}

\noindent Taking the continuous limit of Eq. (\ref{bi}) and utilizing Eqs. (\ref{coord}) and (\ref{mismatch}) we obtain

\begin{eqnarray}
\mathcal{A}({t_{1}}^{d},{t_{2}}^{d})\lefteqn{=iC\xi^{2}\int dk_{o}\int dk_{e}{\int_{0}}^{L} dz e^{-\mathcal{D}\left(\frac{\nu_{o}}{\sigma_{o}}\right)^{2}} e^{-\mathcal{D}\left(\frac{\nu_{e}}{\sigma_{e}}\right)^{2}}}\nonumber \\
& &\times e^{-i(\nu_{e}+\Omega_{e}){t_{2}}^{d}} e^{-i(\nu_{o}+\Omega_{o}){t_{1}}^{d}}e^{-\mathcal{D}\left(\frac{\nu_{o}+\nu{e}}{\sigma_{p}}\right)^{2}}\nonumber \\
& &\times e^{iz\left(u_{e}\nu_{e}+u_{o}\nu_{o}\right)}.\nonumber
\end{eqnarray}

\noindent Where $u_{e}=\frac{1}{U_{p}}-\frac{1}{U_{e}}$ and $u_{o}=\frac{1}{U_{p}}-\frac{1}{U_{o}}$. Note that $dk=\frac{dk}{d\omega}d\omega=\frac{1}{U}d\omega=\frac{1}{U}d\nu$,
so that up to a constant we may switch integration
variables between momentum and frequency. (Actually the $U$'s are frequency dependent, however they do not vary significantly over the bandwidth of the field.) Here we diverge from Kim
and Shih in that we integrate over $\nu_{o}$ and $\nu_{e}$ instead
of $\nu_{-}\equiv \nu_{o}-\nu_{e}$, and $\nu_{p}$ and we do not take
the filter bandwidths to be equivalent. Now

\begin{eqnarray}
\mathcal{A}({t_{1}}^{d},{t_{2}}^{d})\lefteqn{=\frac{iC\xi^{2}}{U_{e}U_{o}}e^{-i(\Omega_{o}{t_{1}}^{d}+\Omega_{e}{t_{2}}^{d})}\int d\nu_{o}\int d\nu_{e}{\int_{0}}^{L} dz} \nonumber \\
& &\times e^{-\mathcal{D}\left(\frac{\nu_{o}}{\sigma_{o}}\right)^{2}} e^{-\mathcal{D}\left(\frac{\nu_{e}}{\sigma_{e}}\right)^{2}}e^{-i\nu_{e}{t_{2}}^{d}} e^{-i\nu_{o}{t_{1}}^{d}} \nonumber \\
& &\times e^{-\mathcal{D}\left(\frac{\nu_{o}+\nu{e}}{\sigma_{p}}\right)^{2}}e^{iz\left(u_{e}\nu_{e}+u_{o}\nu_{o}\right)}. \nonumber
\end{eqnarray}

We have the unitary Fourier transform over $\nu_{o}$, $\nu_{e}$ (up to a factor of $\sqrt{2\pi}$), and the integral over $z$ given by

\begin{eqnarray}
\mathcal{A}({t_{1}}^{d},{t_{2}}^{d})\lefteqn{=\frac{2\pi iC\xi^{2}}{U_{e}U_{o}}e^{-i(\Omega_{o}{t_{1}}^{d}+\Omega_{e}{t_{2}}^{d})}}\nonumber \\
& &\times {\int_{0}}^{L} dz\hat{\mathcal{F}}(\nu_{e}\rightarrow {t_{2}}^{d}) \hat{\mathcal{F}}(\nu_{o}\rightarrow {t_{1}}^{d}) \Upsilon (\nu_{o},\nu_{e},z), \nonumber \\
\Upsilon (\nu_{o},\nu_{e},z)\lefteqn{=e^{-\mathcal{D}\left[\left(\frac{\nu_{o}+\nu{e}}{\sigma_{p}}\right)^{2}+\left(\frac{\nu_{o}}{\sigma_{o}}\right)^{2}+\left(\frac{\nu_{e}}{\sigma_{e}}\right)^{2}\right]}}\nonumber \\
& &\times e^{iz(u_{e}\nu_{e}+u_{o}\nu_{o})}. \label{up}
\end{eqnarray}

\noindent Where we have defined the biphoton kernel $\Upsilon (\nu_{o},\nu_{e},z)$. In the interest of clarity the Fourier transforms have been written as operators

\begin{eqnarray}
\hat{\mathcal{F}}(x\rightarrow y)\equiv \frac{1}{\sqrt{2\pi}}\int^{\infty}_{-\infty}dxe^{-ixy} \nonumber
\end{eqnarray}

\noindent Using $\mathrm{Mathematica}^{\mathrm{TM}}$, we perform these operations in the above order. Note that another ordering will lead to the problem becoming intractable. The result is given by Eq.(\ref{biphoton}). For the sake of simplicity in most of our calculations we will take $\xi=1$.

\subsection*{State Normalization}
The state $|\Psi\rangle $ needs to be normalized. Take again Eq.(\ref{SPDC}) where the integral over $\omega_{p}$ has been performed

\begin{eqnarray}
|\Psi\rangle =C\sum_{k_{e}k_{o}}\int^{L}_{0}dz e^{-\mathcal{D}\left(\frac{\nu_{e}+\nu_{o}}{\sigma_{p}}\right)^{2}}e^{iz\Delta_{k_{e}k_{o}}}\hat{a}^{\dagger}_{e}\hat{a}^{\dagger}_{o}|0\rangle ,\nonumber
\end{eqnarray}

\noindent Now we take the inner product

\begin{eqnarray}
\langle \Psi |\Psi \rangle &=& \frac{4C^{2}}{U_{e}^{2}U_{o}^{2}}\int^{\infty}_{-\infty}\int^{\infty}_{-\infty}d\nu_{e}d\nu_{o}e^{-2\mathcal{D}\left(\frac{\nu_{s}+\nu_{i}}{\sigma_{p}}\right)^{2}}\nonumber \\
&\times &\frac{\sin^{2}\left(\frac{L}{2}\left[\nu_{e}u_{e}+\nu_{o}u_{o}\right]\right)}{\left[\nu_{e}u_{e}+\nu_{o}u_{o}\right]^{2}}.\nonumber
\end{eqnarray}

\noindent Where we have taken the continuous limit and switched integration variables. It is required that $\langle \Psi |\Psi \rangle =1$. We make another change of variables: $p=\nu_{e}+\nu_{o}$ and $q=\nu_{e}u_{e}+\nu_{o}u_{o}$, as long as $U_{e}\neq U_{o}$ this is a well defined one-to-one transformation with a Jacobian $J=\frac{U_{e}U_{o}}{U_{e}-U_{o}}$. This allows us to separate the integral. Thus

\begin{eqnarray}
\frac{4C^{2}}{U_{e}U_{o}(U_{e}-U_{o})}\int_{-\infty}^{\infty}&dp& e^{-2\mathcal{D}\left(\frac{p}{\sigma_{p}}\right)^{2}}\int_{-\infty}^{\infty}dq\frac{\sin^{2}\left(\frac{L}{2}q\right)}{q^{2}} \nonumber \\
&=&\frac{2C^{2}}{U_{e}U_{o}(U_{e}-U_{o})}\frac{\pi^{\frac{3}{2}} L\sigma_{p}}{\sqrt{2\mathcal{D}}}.\nonumber
\end{eqnarray}

\noindent So

\begin{eqnarray}
C=\left(\frac{\mathcal{D}}{2}\right)^{\frac{1}{4}}\frac{\sqrt{U_{e}U_{o}(U_{e}-U_{o})}}{\pi^{\frac{3}{4}}\sqrt{L\sigma_{p}}}.\nonumber
\end{eqnarray}

\noindent This factor is added onto Eq. (\ref{p2t}). Using a similar procedure we obtain for the continuous wave case

\begin{eqnarray}
C_{\mathrm{cw}}=\sqrt{\frac{U_{e}U_{o}(U_{e}-U_{o})}{2\pi L}}, \nonumber
\end{eqnarray}   

\noindent which we append to Eq. (\ref{cw}). 

\section*{Appendix B: Calculation of the two photon absorption probability for N00N states}

Noting that Eq. (\ref{S}) represents a unitary transform back from time space into frequency space we may write the $\mathcal{Z}$ function as

\begin{eqnarray}
\mathcal{Z}(\omega_{o},\omega_{e})=2\pi\hat{\mathcal{F}}({t_{1}}^{d}\rightarrow \omega_{o})^{*}\hat{\mathcal{F}}({t_{2}}^{d}\rightarrow \omega_{e})^{*}A({t_{1}}^{d},{t_{2}}^{d}).\nonumber
\end{eqnarray}

\noindent Therefore

\begin{widetext}
\begin{eqnarray}
\mathcal{Z}(\omega_{o},\omega_{e})\lefteqn{=2\pi\hat{\mathcal{F}}({t_{1}}^{d}\rightarrow \omega_{o})^{*}\hat{\mathcal{F}}({t_{2}}^{d}\rightarrow \omega_{e})^{*}A({t_{1}}^{d},{t_{2}}^{d})} \nonumber \\
\lefteqn{=2\pi\hat{\mathcal{F}}({t_{1}}^{d}\rightarrow \omega_{o})^{*}\hat{\mathcal{F}}({t_{2}}^{d}\rightarrow \omega_{e})^{*}\mathcal{A}({t_{1}}^{d},{t_{2}}^{d})+2\pi\hat{\mathcal{F}}({t_{1}}^{d}\rightarrow \omega_{o})^{*}\hat{\mathcal{F}}({t_{2}}^{d}\rightarrow \omega_{e})^{*}\mathcal{A}({t_{2}}^{d},{t_{1}}^{d})} \nonumber \\
\lefteqn{=C'\hat{\mathcal{F}}({t_{1}}^{d}\rightarrow \omega_{o})^{*}\hat{\mathcal{F}}({t_{2}}^{d}\rightarrow \omega_{e})^{*}e^{-i(\Omega_{o}{t_{1}}^{d}+\Omega_{e}{t_{2}}^{d})}{\int_{0}}^{L} dz\hat{\mathcal{F}}(\nu_{e}\rightarrow {t_{2}}^{d}) \hat{\mathcal{F}}(\nu_{o}\rightarrow {t_{1}}^{d}) \Upsilon (\nu_{o},\nu_{e},z)}\nonumber\\
& &+C'\hat{\mathcal{F}}({t_{1}}^{d}\rightarrow \omega_{o})^{*}\hat{\mathcal{F}}({t_{2}}^{d}\rightarrow \omega_{e})^{*}e^{-i(\Omega_{o}{t_{2}}^{d}+\Omega_{e}{t_{1}}^{d})}{\int_{0}}^{L} dz\hat{\mathcal{F}}(\nu_{e}\rightarrow {t_{1}}^{d}) \hat{\mathcal{F}}(\nu_{o}\rightarrow {t_{2}}^{d}) \Upsilon (\nu_{o},\nu_{e},z) \nonumber \\
\lefteqn{=C'{\int_{0}}^{L} dz\hat{\mathcal{F}}({t_{2}}^{d}\rightarrow \nu_{e})^{*}\hat{\mathcal{F}}(\nu_{e}\rightarrow {t_{2}}^{d}) \hat{\mathcal{F}}({t_{1}}^{d}\rightarrow \nu_{o})^{*}\hat{\mathcal{F}}(\nu_{o}\rightarrow {t_{1}}^{d}) \Upsilon (\nu_{o},\nu_{e},z)} \nonumber \\
& &+C'{\int_{0}}^{L} dz\hat{\mathcal{F}}({t_{1}}^{d}\rightarrow \nu_{o})^{*}\hat{\mathcal{F}}(\nu_{e}\rightarrow {t_{1}}^{d}) \hat{\mathcal{F}}({t_{2}}^{d}\rightarrow \nu_{e})^{*}\hat{\mathcal{F}}(\nu_{o}\rightarrow {t_{2}}^{d}) \Upsilon (\nu_{o},\nu_{e},z) \nonumber \\
\lefteqn{=C'{\int_{0}}^{L} dz\Upsilon (\nu_{o},\nu_{e},z)+C'{\int_{0}}^{L} dz\Upsilon (\nu_{e},\nu_{o},z).}\label{speccor}
\end{eqnarray}
\end{widetext}

\noindent In the second equality we have made a substitution
using Eq. (\ref{bi}). In the third equality Eq.(\ref{up}) was used. In the fourth equality the overall phases have combined with the Fourier transform operators to change their output variables. So the Fourier transforms from the biphoton amplitude exactly cancel with the Fourier transforms from the definition of the spectral correlation function, leaving a relatively simple result. Note that $\Upsilon(\nu_{o},\nu_{e},z)\neq\Upsilon(\nu_{e},\nu_{o},z)$. And

\begin{eqnarray}
C'=2\pi i(8\pi\mathcal{D})^{\frac{1}{4}}\sqrt{\frac{U_{2}}{L\sigma_{p}}}\nonumber
\end{eqnarray}

Where $U_{2}$ is defined in the main text. Returning to Eq. (\ref{p2}) we make the assumption that $g(\omega)\approx
g(\frac{1}{2}\Omega_{p})$. To see why, we need to investigate the definition of the atomic response function, Eq. (3.8) from Mollow \cite{Mollow}  

\begin{eqnarray}
g(\omega)=\mu\sum_{j}p_{fj}p_{j0}\frac{1}{\omega -\omega_{j}+\frac{1}{2}\kappa_{j}}.\nonumber
\end{eqnarray}

\noindent Where $\mu$ is a constant representative of the absorber, $j$ labels the possible intermediate (virtual) levels, $p_{j0}$ and $p_{fj}$ are the momentum matrix elements of the electron making a transition between the initial and $j$th states, and the $j$th and final states respectively. $\kappa_{j}$ is the linewidth of the intermediate level. Each term in the series represents the two photon absorption process proceeding via a different intermediate level transition. We assume that the central frequency of the field is one half the resonant frequency of the final state. Note that each term of $g$ is highly peaked around the characteristic frequency of the level, which will be far detuned from the central frequency of the field. Unless the bandwidth of the field is on the order of the transition frequency, the value of the atomic response function will not change much as $\omega $ is varied over the bandwidth of the incident light. Therefore our assumption is justified. Note that this is the \textit{only} assumption we make about bandwidths in this paper. It is worth noting that there has been much work done towards engineering materials which have large two photon cross sections \cite{mat1,mat2}, so as light sources are improving, so are absorbers.   

So, using Eq. (\ref{speccor}) and Eq. (\ref{p2}) we have

\begin{eqnarray}
P_{2}&=&\left|C'g\left(\frac{1}{2}\Omega_{p}\right)\int^{\infty}_{-\infty}d\nu\int^{\infty}_{-\infty}dz\right.\nonumber\\
& &\times\left[\Upsilon(\nu_{f}-\nu,\nu,z)+\Upsilon(\nu,\nu_{f}-\nu,z)\right]|^{2}.\label{p21}
\end{eqnarray}

The importance of the atomic response function being
separated from the integral should be noted. Essentially this means
that the spectral properties of the light may be considered apart
from the structure of the absorber. The remaining integral is not difficult to perform 

\begin{eqnarray}
{\int_{0}}^{L}&dz&\int^{\infty}_{-\infty}d\nu\Upsilon (\nu_{f}-\nu,\nu,z)\nonumber \\
&=&\frac{\pi}{U_{2}}e^{-\mathcal{D}{\nu_{p}}^{2}\left(\frac{(u_{e}+U_{2})^{2}}{\sigma_{e}^{2}U_{2}^{2}}+\frac{u_{e}^{2}}{\sigma_{o}^{2}U_{2}^{2}}+\frac{1}{\sigma_{p}^{2}}\right)}\nonumber \\
&&\times\left[\mathrm{Erf}\left(\mathcal{E}\nu_{p}\right) -\mathrm{Erf}\left(\mathcal{E}\nu_{p}-\mathcal{L}\right)\right]. \nonumber
\end{eqnarray}

\noindent Where $U_{2}$, $u_{e}$, $\mathcal{E}$, and $\mathcal{L}$ are defined in the main text in Eq. (\ref{el}). The integral over $\Upsilon(\nu,\nu_{f}-\nu,z)$ differs only by a constant factor. 

However, the utility of this expression is limited. It only describes the probability of a single frequency from the pump beam being absorbed. Furthermore it implies that this frequency will be on resonant with the final level. Realistically, all frequencies will have a chance to be absorbed, and only $\nu_{f}=0$ will be resonant. To correct for this we can average the function over a lorentzian line shape with its peak at $\nu_{f}=0$ and a height of 1. The full equation is given in the main text by Eq. (\ref{p2t}).

\section*{Appendix C: The Fair Comparison Coherent State}

Again, our $|2::0\rangle $ state is produced from a photon pair from a type II spontaneous parametric down conversion event, incident on two filters and a symmetric beam splitter. We can examine $|\Psi\rangle $, Eq. (\ref{SPDC}), and see that \--- given the assumptions in Appendix A \--- this state has a spectral profile (the relative probability amplitude of the photons having frequencies $\nu_{e}$ and $\nu_{o}$) given by 

\begin{eqnarray}
F_{|2::0\rangle}(\nu_{e},\nu_{o})=C\int^{L}_{0}dze^{-\mathcal{D}\left(\frac{\nu_{e}+\nu_{o}}{\sigma_{p}}\right)^2}e^{iz\left(\frac{\nu_{e}+\nu_{o}}{U_{p}}-\frac{\nu_{e}}{U_{e}}-\frac{\nu_{o}}{U_{o}}\right)}.\nonumber
\end{eqnarray}

\noindent $|\Psi\rangle $ represents two entangled down conversion modes. To remove the non-classical nature of this light we project out one of the modes. We arbitrarily choose the ordinary mode. The remaining state will have the spectral profile we desire for one mode of our fair coherent state ($F_{\alpha}$), that is, 

\begin{eqnarray}
\int^{\infty}_{-\infty}d\nu_{o}\langle 1_{\nu_{o}}|\Psi\rangle &=&\sqrt{\frac{\pi\sigma_{p}^{2}}{\mathcal{D}}}\sum_{k'}F_{\alpha ,z}(\nu_{e})\hat{a}^{\dagger}_{ek'}|0\rangle_{e},\nonumber\\
F_{\alpha}(\nu_{e})&=&\int^{L}_{0}dze^{-iz\nu_{e}(u_{o}-u_{e})}e^{-\frac{u_{o}^{2}z^{2}\sigma_{p}^{2}}{4\mathcal{D}}},\nonumber \\ \nonumber
\end{eqnarray}

\noindent where $u_{e}=1/U_{p}-1/U_{e}$ and $u_{o}=1/U_{p}-1/U_{o}$. This will represent one mode of the ``fair comparison" coherent light. The normalization will be discussed shortly. The other mode will have a separate, but identical, spectral profile. Both modes will be mixed with a symmetric beam splitter.

In order to assimilate this spectral information into a coherent state we identify the spectral profile with $\alpha $ thus: $\alpha_{k}=F_{\alpha_{k}}(\nu_{k})$.

The beam splitter transformation operating on the displacement operators which create the two mode coherent state, $|\alpha\rangle_{1}|\beta\rangle_{2}$ produces an output of $|\frac{\alpha+i\beta}{\sqrt{2}}\rangle_{3}|\frac{i\alpha+\beta}{\sqrt{2}}\rangle_{4}$ . Where $1$ and $2$ label the input modes and $3$ and $4$ label the output modes.

So now we can write the temporal correlation function for the fair coherent state

\begin{widetext}
\begin{eqnarray}
G^{(2)}_{\alpha}({t'}_{1}^{d},{t'}_{2}^{d},t_{1}^{d},t_{2}^{d})&=&\xi^{4}\bigotimes_{\nu}\left\langle\frac{\alpha+i\beta}{\sqrt{2}}\gamma_{3}\right|_{3\nu}\left\langle\frac{i\alpha +\beta}{\sqrt{2}}\gamma_{4}\right|_{4\nu}\sum_{\nu_{a}} \hat{a}_{3\nu_{a}}^{\dagger}e^{i(\nu_{a}+\Omega){t'}^{d}_{1}}e^{-\mathcal{D}\left(\frac{\nu_{a}}{\sigma '}\right)^{2}}\sum_{\nu_{b}} \hat{a}_{4\nu_{b}}^{\dagger}e^{i(\nu_{b}+\Omega){t'}^{d}_{2}}e^{-\mathcal{D}\left(\frac{\nu_{b}}{\sigma '}\right)^{2}}\nonumber\\
& &\times\sum_{\nu_{c}}\hat{a}_{3\nu_{c}} e^{-i(\nu_{c}+\Omega){t}^{d}_{1}}e^{-\mathcal{D}\left(\frac{\nu_{c}}{\sigma '}\right)^{2}}\sum_{\nu_{d}}\hat{a}_{4\nu_{d}} e^{-i(\nu_{d}+\Omega){t}^{d}_{2}}e^{-\mathcal{D}\left(\frac{\nu_{d}}{\sigma '}\right)^{2}}\bigotimes_{\nu}\left|\frac{\alpha+i\beta}{\sqrt{2}}\gamma_{3}\right\rangle_{3\nu}\left|\frac{i\alpha +\beta}{\sqrt{2}}\gamma_{4}\right\rangle_{4\nu}\nonumber \\
&=&\xi^{4}\pi^{2}|\gamma_{3}|^{2}|\gamma_{4}|^{2}e^{i\Omega({t'}^{d}_{1}+{t'}^{d}_{2}-{t}^{d}_{1}-{t}^{d}_{2})}
\hat{\mathcal{F}}(\nu_{a}\rightarrow{t'}^{d}_{1})^{*}\hat{\mathcal{F}}(\nu_{b}\rightarrow{t'}^{d}_{2})^{*}\hat{\mathcal{F}}(\nu_{c}\rightarrow{t}^{d}_{1})\hat{\mathcal{F}}(\nu_{d}\rightarrow{t}^{d}_{2})\nonumber\\
& &\times\left[F^{*}_{\alpha}(\nu_{a})-iF^{*}_{\beta}(\nu_{a})\right]\left[-iF^{*}_{\alpha}(\nu_{b})+F^{*}_{\beta}(\nu_{b})\right]
\left[F_{\alpha}(\nu_{c})+iF_{\beta}(\nu_{c})\right]\left[iF_{\alpha}(\nu_{d})+F_{\beta}(\nu_{d})\right].\nonumber
\end{eqnarray}
\end{widetext}

\noindent Where the factors $\gamma $ are the normalizations in the indexed modes. The central frequencies ($\Omega$) have been taken to be the same. The filters in the arms $\sigma_{e}=\sigma_{o}$ have been removed in order to make the calculation which follows mathematically tractable. The frequencies have been labeled $a$,$b$,$c$, and $d$. As before $\xi$ represents the constants associated with the electric field operators. We shall eventually take the two two photon absorption probabilities in ratio, the $\xi$'s will cancel exactly. In light of this we simply set $\xi=1$. We need to normalize such that the information about the intensity of the coherent light in each arm is contained in the $\gamma $'s.

\begin{eqnarray}
I&=&\int^{\infty}_{-\infty}dt\bigotimes_{\omega}\left\langle\gamma\alpha_{\omega}\right|\sum_{\omega '}\hat{a}^{\dagger}_{\omega '}e^{i\omega 't}\nonumber\\
& &\times\sum_{\omega ''}\hat{a}_{\omega ''}e^{-i\omega ''t}\bigotimes_{\omega}\left|\gamma\alpha_{\omega}\right\rangle\nonumber\\
&=&|\gamma |^{2}\int^{\infty}_{-\infty}d\omega '\int^{\infty}_{-\infty}d\omega ''\delta(\omega '-\omega '')F_{\alpha}^{*}(\omega ')F_{\alpha}(\omega '') \nonumber\\
&=&|\gamma |^{2}\int^{\infty}_{-\infty}d\omega '|F_{\alpha}(\omega ')|^{2}\nonumber\\
&=&|\gamma |^{2}\int^{\infty}_{-\infty}d\nu '\int^{L}_{0}dz_{1}\int^{L}_{0}dz_{2}\nonumber\\
& &\times e^{i(z_{1}-z_{2})\nu '(u_{o}-u_{e})}e^{-\frac{u_{o}^{2}\sigma_{p}^{2}}{4\mathcal{D}}(z_{1}^{2}+z_{2}^{2})}\nonumber\\
&=&\frac{|\gamma|^{2}\sqrt{2\pi}}{u_{o}-u_{e}}\int^{L}_{0}dz_{1}\int^{L}_{0}dz_{2}\delta (z_{1}-z_{2})e^{-\frac{u_{o}^{2}\sigma_{p}^{2}}{4\mathcal{D}}(z_{1}^{2}+z_{2}^{2})}.\nonumber
\end{eqnarray} 

\noindent The double integral is tractable yielding the result

\begin{eqnarray}
\gamma =\left(\frac{I^{2}(u_{o}-u_{e})^{2}u_{o}^{2}\sigma_{p}^{2}}{\mathcal{D}\pi^{2}}\right)^{\frac{1}{4}}\mathrm{Erf}\left(\frac{Lu_{o}\sigma_{p}}{\sqrt{2\mathcal{D}}}\right)^{-\frac{1}{2}}.\nonumber
\end{eqnarray} 

Using the same procedure as in Eq. (\ref{speccor}) we obtain the spectral correlation function

\begin{eqnarray}
S^{(2)}&=&2\pi^{2}|\gamma_{3}|^{2}|\gamma_{4}|^{2}\left[F^{*}_{\alpha}({\nu '}_{1})-iF^{*}_{\beta}({\nu '}_{1})\right]\nonumber\\
& &\times\left[F_{\alpha}(\nu_{1})+iF_{\beta}(\nu_{1})\right]\left[iF_{\alpha}(\nu_{2})+F_{\beta}(\nu_{2})\right]\nonumber\\
& &\times \left[F^{*}_{\beta}({\nu '}_{2})-iF^{*}_{\alpha}({\nu '}_{2})\right].\nonumber
\end{eqnarray}

\noindent Now we can write the two-photon absorption probability Eq. (\ref{p2}) as

\begin{eqnarray}
P_{2}^{\alpha}&=&\pi^{2}|\gamma|^{4}\left|\int^{\infty}_{-\infty}d\nu F_{\alpha}(\nu_{f}-\nu)F_{\alpha}(\nu)\right|^{2}\nonumber\\
&=&\pi^{2}|\gamma|^{4}\left|\int^{\infty}_{-\infty}\frac{d\nu_{u}}{u_{o}-u_{e}}\int^{L}_{0}dz_{1}\int^{L}_{0}dz_{2}e^{-iz_{2}\nu_{f}(u_{o}-u_{e})}\right.\nonumber\\
& &\times \left. e^{-i(z_{1}-z_{2})\nu_{u}}e^{\frac{u_{o}^{2}\sigma_{p}^{2}}{4\mathcal{D}}(z_{1}+z_{2})}\right|^{2}\nonumber\\
&=&\frac{\pi^{2}|\gamma|^{4}}{(u_{o}-u_{e})^{2}}\left|\int^{L}_{0}dz_{1}\int^{L}_{0}dz_{2}e^{-iz_{2}\nu_{f}(u_{o}-u_{e})}\right.\nonumber\\
& &\times\left.\delta(z_{1}-z_{2})e^{\frac{u_{o}^{2}\sigma_{p}^{2}}{4\mathcal{D}}(z_{1}+z_{2})}\right|^{2}\nonumber\\
&=&\frac{\pi^{3}\mathcal{D}|\gamma|^{4}}{2u_{o}^{2}\sigma_{p}^{2}(u_{o}-u_{e})^{2}}e^{-\frac{\mathcal{D}(u_{o}-u_{e})^{2}\nu_{f}^{2}}{u_{o}^{2}\sigma_{p}^{2}}}\nonumber\\
& &\times|\mathrm{Erf}(\mathcal{E}_{\alpha}\nu_{f})-\mathrm{Erf}(\mathcal{E}_{\alpha}\nu_{f}+\mathcal{L}_{\alpha})|^{2}.\nonumber
\end{eqnarray}

\noindent Where $\mathcal{E}_{\alpha}$ and $\mathcal{L}_{\alpha}$ are defined in the main text. In the interest of simplicity the initial coherent states have been taken to be equivalent. As in Appendix B we must now take the integral of the above expression, times a lorentzian line shape, over $\nu_{f}$. This new function, $\widetilde{P}_{2}^{\alpha}$, is given in the main text by Eq. (\ref{p2coh}).


\begin{thebibliography}{99}
\bibitem{Boto}Agedi N. Boto, Pieter Kok, Daniel S. Abrams, Samuel L. Braunstein, Colin P. Williams, and Jonathan P. Dowling, ``Quantum Interferometric Optical Lithography: Exploiting Entanglement to Beat the Diffraction Limit"., \textit{Phys. Rev. Lett.} \textbf{85}, 2733 (2000).
\bibitem{lith}Robert W. Boyd, Sean J. Bently, ``Recent progress in quantum and nonlinear optical lithography", \textit{J. Mod. Optic.} \textbf{53}, 713 (2006).
\bibitem{Dowling}Hwang Lee, Pieter Kok, Jonathan P. Dowling, ``Quantum Imaging and Metrology", \textit{arXiv:quant-ph/0306113v1} (2008).
\bibitem{met}Vittorio Giovannetti, Seth Lloyd, and Lorenzo Maccone, ``Quantum Metrology", \textit{Phys. Rev. Lett.} \textbf{96}, 010401 (2006).
\bibitem{Gea}J. Gea-Banacloche, ``Two-Photon Absorption of Nonclassical Light", \textit{Phys. Rev. Lett.} \textbf{62}, 1603 (1989).
\bibitem{line}Juha Javanainen, and Phillip L. Gould, ``Linear Intensity Dependence of a Two-Photon Transition Rate", \textit{Phys. Rev. A} \textbf{41}, 5088 (1990).
\bibitem{proof}N. Ph. Georgiades, E.S. Polzik, K. Edamatsu, and H.J. Kimble, ``Nonclassical Excitation for Atoms in a Squeezed Vacuum", \textit{Phys. Rev. Lett.} \textbf{75}, 3426 (1995).
\bibitem{sakurai}J. J. Sakurai, \emph{Modern Quantum Mechanics p.333}, (Addison-Wesley, Reading, MA, 1994).
\bibitem{Agarwal}G. S. Agarwal, ``Field-Correlation Effects in Multiphoton Absorption Processes", \textit{Phys. Rev. A} \textbf{1}, 1445 (1970).
\bibitem{Tsang}Mankei Tsang, ``On the Relationship between Resolution Enhancement and Multiphoton Absorption Rate in Quantum Lithography", \textit{Phys. Rev A} \textbf{75}, 043813 (2007).
\bibitem{Harris}S.E. Harris, ``Chirp and Compress: Toward Single-Cycle Biphotons", \textit{Phys. Rev. Lett.} \textbf{98}, 063602 (2007).
\bibitem{Du}Shengwang Du, Jianming Wen, Chinmay Belthangady, ``Temporally shaping biphoton wave packets with periodically modulated driving fields", \textit{Phys. Rev. A} \textbf{79}, 043811 (2009).
\bibitem{new}Dmitry V. Strekalov, Matthew C. Stowe, Maria V. Chekhova
and Jonathan P. Dowling, ``Two-photon processes in faint biphoton fields", \textit{J. Mod. Optic.} \textbf{49}, 2349 (2002).
\bibitem{dayan1}Barak Dayan, Avi Pe'er, Asher A. Friesem, and Yaron Silberberg, ``Two Photon Absorption and Coherent Control with Broadband Down-Converted Light", \textit{Phys. Rev. Lett.} \textbf{93}, 023005 (2004).
\bibitem{Teich}Adel Joobeur, Bahaa E. A. Saleh, Todd S. Larchuk, and Malvin C. Teich, ``Coherence properties of entangled light beams generated by parametric down-conversion:
Theory and experiment", \textit{Phys. Rev. A} \textbf{53}, 4360 (1996).
\bibitem{torres}M. Hendrych, Xiaojuan Shi, A. Valencia, Juan P. Torres, ``Broadening the bandwidth of entangled photons: A step towards the generation
of extremely short biphotons", \textit{Phys. Rev. A} \textbf{29}, 023817 (2009).
\bibitem{dayan2}Barak Dayan, ``Theory of two-photon interactions with broadband down-converted light and entangled photons", \textit{Phys. Rev. A} \textbf{76}, 043813 (2007). 
\bibitem{Rubin}Morton H. Rubin, David N. Klyshko, Y. H. Shih, and A. V. Sergienko, ``Theory of two-photon entanglement in type-II optical parametric down-conversion", \textit{Phys. Rev. A} \textbf{50}, 5122 (1994).
\bibitem{HOM}C. K. Hong, Z. Y. Ou, and L. Mandel, ``Measurement of Subpicosecond Time Intervals between Two Photons by Interference", \textit{Phys. Rev. Lett.} \textbf{59}, 2044 (1987).
\bibitem{con1}D.S. Elliott, M.W. Hamilton, K. Arnett, and S.J. Smith, ``Correlation effects of a phase-diffusing field on two-photon absorption", \textit{Phys. Rev. A} \textbf{32}, 887 (1985).
\bibitem{con2}Ce Chen, D.S. Elliott, and M.W. Hamilton, ``Two-Photon Absorption from the Real Gaussian Field", \textit{Phys. Rev. A} \textbf{68}, 3531 (1992).
\bibitem{Mollow}B. R. Mollow, ``Two-Photon Absorption and Field Correlation Functions", \textit{Phys. Rev.} \textbf{175}, 1555 (1968).
\bibitem{Glauber}Roy J. Glauber, ``The Quantum Theory of Optical Coherence", \textit{Phys. Rev.} \textbf{130}, 2529 (1963).
\bibitem{Keller}Timothy E. Keller and Morton H. Rubin, ``Theory of two-photon entanglement for spontaneous parametric down-conversion driven by a narrow pump pulse", \textit{Phys. Rev. A} \textbf{56}, 1534 (1997).
\bibitem{Kim}Yoon-Ho Kim, Vincenzo Berardi, Maria V. Chekhova, Augusto Garuccio, and Yanhua Shih, ``Temporal indistinguishability and quantum interference", \textit{Phys. Rev. A} \textbf{62}, 043820 (2000).
\bibitem{thesis}Yoon-Ho Kim, \textit{Two-Photon Quantum Entanglement}, (Thesis Dissertation, University of Maryland, 2001).
\bibitem{shih}Yanhua Shih, ``Entangled biphoton source - property and
preparation" \textit{Rep. Prog. Phys.} \textbf{66}, 1009 (2003).
\bibitem{bbo}D. N. Nikogosyan, ``Beta Barium Borate", \textit{Applied Physics A} \textbf{52}, 359 (1991).
\bibitem{eff}Christian Kurtsiefer, Markus Oberparleiter, and Harald Weinfurter, ``High-efficiency entangled photon pair collection in type-II parametric fluorescence", \textit{Phys. Rev. A} \textbf{64}, 023802 (2001).
\bibitem{trans}Hong-Bing Fei, Bradley M. Jost, Sandu Popescu, Bahaa E.A. Saleh, and Malvin C. Teich, ``Entanglement-Induced Two-Photon Transparency", \textit{Phys. Rev. Lett.} \textbf{78}, 1679 (1997).
\bibitem{mat1}Michael R. Harpham, \"Ozg\"un S\"uzer, Chang-Qi Ma, Peter B\"auerle, and Theodore Goodson III ``Thiophene Dendrimers as Entangled Photon Sensor Materials", \textit{J. Am. Chem. Soc.} \textbf{131}, 973 (2009).
\bibitem{mat2}Dong-Ik Lee, and Theodore Goodson III, ``Entangled Photon Absorption in an Organic Porphyrin Dendrimer", \textit{J. Phys. Chem. B} \textbf{110}, 25582 (2006).
\end{thebibliography}
\end{document}